\documentclass[11pt,a4paper]{article}
\usepackage{jheppub,slashed}
\def\be{\begin{equation}}
\def\ee{\end{equation}}
\def\ba{\begin{array}}
\def\ea{\end{array}}

\newcommand{\bea}{\begin{eqnarray}}
\newcommand{\eea}{\end{eqnarray}}

\def\K{K{\"a}hler}
\newcommand{\rf}[1]{(\ref{#1})}

\def\Re{\mathop{\rm Re}\nolimits}
\def\Im{\mathop{\rm Im}\nolimits}
\newcommand{\hc}{{\rm h.c.}}
\newcommand{\ft}[2]{{\textstyle\frac{#1}{#2}}}
\def\rmi{{\rm i}}
\def\rmd{{\rm d}}

\newcommand{\bbox}{\lower.2ex\hbox{$\Box$}}

\title{On the Supersymmetric Completion of $R+R^2$ Gravity and Cosmology}

\author [a,b]{Sergio Ferrara,} \author  [c]{Renata  Kallosh,}   \author[d]{Antoine Van Proeyen,}
\affiliation[a]{\sl Physics Department, Theory Unit, CERN, CH 1211, Geneva 23, Switzerland}
\affiliation[b]{\sl $^{2}$ INFN - Laboratori Nazionali di Frascati, Via Enrico Fermi 40, I-00044 Frascati,
Italy~\footnote{On leave of absence from Department of Physics and Astronomy, University of California Los Angeles, CA 90095-1547 USA}
}
\affiliation[c]{\sl Stanford Institute for Theoretical Physics and Department of Physics, Stanford University,\\ Stanford, CA 94305-4060, USA}
\affiliation[d]{\sl Instituut voor Theoretische Fysica, KU Leuven,\\ Celestijnenlaan 200D, B-3001 Leuven,
Belgium}\emailAdd{sergio.ferrara@cern.ch}
\emailAdd{kallosh@stanford.edu}  \emailAdd{Antoine.VanProeyen@fys.kuleuven.be}

\abstract{We revisit and clarify the  supersymmetric versions of $R+ R^2$ gravity, in view of the renewed interest to these models in cosmology.
We emphasize that the content of the dual standard supergravity theory in the old minimal formulation necessarily includes two massive chiral multiplets,
that we call the inflaton and the goldstino.
We point out that the presence of these multiplets is model independent in the old minimal formulation and therefore any theory
that contains a single chiral multiplet fails to be a supersymmetric generalization of the $R+R^2$ gravity. The supergravity interactions of the two chiral multiplets are encoded in a superpotential mass term and an arbitrary K{\"a}hler potential for the goldstino multiplet. The implication for cosmology of the supersymmetric  $R+R^2$ gravity
is also discussed.
}

\begin{document}
\begin{flushright}
CERN-PH-TH/2013-220
\end{flushright}
 \maketitle
\section{Introduction}
We discuss higher curvature theories of supergravity which can be
considered as the supersymmetric version of the $R+ R^2$ bosonic theory.
The supersymmetric extension depends on the off-shell formulations of
the supergravity. For minimal formulation with six bosonic degrees of
freedom for the auxiliary fields, the $R+ R^2$ theory is dual to a
standard (two derivative) supergravity. In the old minimal case the
linearized analysis \cite{Ferrara:1978rk} reveals two massive chiral
multiplets, with the same mass, one of which  contains the scalaron,
i.e. the gravitational degree of freedom of the bosonic theory. We will
call the first chiral multiplet $\Phi $ an ``inflaton multiplet'' and
the second chiral multiplet $S$, a ``goldstino multiplet", which will be
explained in the context of inflationary evolution.

The full non-linear theory of the  $R+ R^2$ and generalizations thereof
in old minimal supergravity was found in \cite{Cecotti:1987sa} and was
shown to reduce to a standard two-derivative supergravity of two chiral
superfields. In the new minimal supergravity full non-linear theory of
the  $R+ R^2$ was found in \cite{Cecotti:1987qe} and was shown to be
given by a standard two-derivative supergravity interacting with a
massive vector multiplet. However, while the new minimal extension
appears to be unique some freedom exists in the old minimal one. This
freedom manifests in an arbitrary \K\, function for the goldstino chiral
multiplet whose $\theta=0$ component is the sgoldstino. The name of this
second multiplet comes from the observation that during inflation the
auxiliary field $F_S=\frac{\partial W}{\partial S}$ is not vanishing and
is responsible for supersymmetry breaking during inflation. In absence
of this \K\, function one of the two chiral multiplets is no longer
dynamical and the theory is not any longer the supersymmetrization of
the $R+ R^2$ gravity.

The paper is organized as follows. In Sec. \ref{ss:RR2sg} we review $R+R^2$ supergravity and its dual standard
two-derivative supergravity description as considered in \cite{Cecotti:1987sa}.
In the same section we revisit what in the literature was called ``$F(R)$ supergravity'' \cite{Ketov:2009sq,Ketov:2010qz}.
This is the situation where only one chiral multiplet is dynamical.
In this case the supergravity interaction of this chiral multiplet has a fixed \K\, potential but an arbitrary superpotential.
In Sec. \ref{ss:compFR} we present the component expression of ``$F(R)$ supergravity''.
The resulting action shows that while the vector auxiliary field $A_\mu$ appears algebraically and can be eliminated using its algebraic equations of motion,
the auxiliary field $X$ appears covered by a derivative and is therefore propagating.
This completely agrees with the linearized analysis of  \cite{Ferrara:1978rk}.
Note that to confine the analysis to a sector where $X$ only appears algebraically, gives an inconsistent result,
as already noticed in  \cite{Gates:2009hu}.
Our analysis provides a detailed explanation why ``$F(R)$ supergravity''   \cite{Ketov:2009sq,Ketov:2010qz}
has only terms linear in bosonic curvature $R$, which is important for the cosmological applications.
 Sec. \ref{ss:discussion} provides a discussion of our findings.
In an Appendix some elements of a conformal tensor calculus useful for understanding our results in components are given.

\section{ \texorpdfstring{$R+R^2$}{R+R2} supergravity}
\label{ss:RR2sg}
\subsection{Manifestly superconformal action}

In the old minimal supergravity the most general $R+R^2$ theory is given
by a Lagrangian of the form \cite{Cecotti:1987sa}
\begin{equation}
{\cal L}= -\left[ S_0 \cdot  \bar S_0\right] _D +\left[ S_0 \cdot  \bar S_0 \cdot   h\left({{\cal R}\over S_0}, {\bar {\cal R}\over \bar S_0}\right) \right] _D
+\left[ S_0^3 \cdot   W \left({{\cal R}\over S_0}\right)\right] _F\,.
\label{Sergio}
\end{equation}
This action has a manifest superconformal symmetry, and we write it
using notations of superconformal calculus, reviewed in
\cite{Freedman:2012zz}. Old minimal supergravity uses a chiral
compensating multiplet $S_0$ of Weyl weight~1. Another chiral multiplet,
a curvature multiplet ${\cal R}$ of weight~1, is defined as
\cite{Kugo:1983mv}
\begin{equation}
  {\cal R}= (S_0)^{-1} \cdot  \Sigma (S_0)=(S_0)^{-1} \cdot  T(\bar S_0)\,,
 \label{defRchiral}
\end{equation}
where the operation $\Sigma $ or $T$ is defined in (\ref{SigmaT}). The
multiplication of chiral multiplets leads to new chiral multiplets, and
$[\ldots]_F$ denotes the `$F$' action formula on a chiral multiplet of
weight~3. On the other hand, $[\ldots]_D$ indicate the $D$ action
formula on a real multiplet of weight~2. We use the symbol $\cdot$ for
products to distinguish the use of curly brackets as arguments of a
function from a multiplication by an expression in brackets.

The function $h$ is a real function, and $W$ is holomorphic. In order to
have an $R+R^2$ supergravity it is essential that
 $h_{R \bar R}\equiv {\partial^2 h\over \partial {\cal R} \partial \bar {\cal R} }\neq 0$. In fact, in absence of this term, as we will show later, the theory widely considered in the literature \cite{Ketov:2009sq,Ketov:2010qz} is no longer a supersymmetric extension of the
$R+R^2$ theory. In particular, it does not contain two chiral multiplets, as required by a linearized analysis of \cite{Ferrara:1978rk}. In fact, when $h_{R \bar R}= 0$ the goldstino multiplet is no longer dynamical.

\subsection{The dual to $R+R^2$ supergravity}
The dual theory  is obtained by introducing two chiral superfields, $S$ and $\Phi $, both of weight~0, so that  \rf{Sergio} can be written as
\begin{equation}
{\cal L}= -\left[ S_0 \cdot  \bar S_0\right] _D + \left[S_0 \cdot  \bar S_0 \cdot  h\left(S, {\bar S}\right)  \right] _D
 +\left[ S_0^3 \cdot  \left( W (S)   -3\,\Phi \cdot  \left({{\cal R}\over S_0}-S \right) \right)  \right] _F\,.
\label{Sergio1}
\end{equation}
We also write the superpotential term $W$ as follows
\begin{equation}
W(S)= S\cdot g(S)+\lambda\,,
\end{equation}
where $\lambda $ is a constant. By using the lemma, eq. (11) of
\cite{Cecotti:1987sa}, which is also explained in  (\ref{thmDtoF}),
\begin{equation}
\left[ \Phi \cdot  {\cal R} \cdot  S_0^2\right] _F = \left[S_0 \cdot  \bar S_0 \cdot (\Phi +\bar \Phi)  \right] _D + \rm {tot. der}
\label{lemma}\end{equation}
we obtain
\begin{equation}
{\cal L}= \left[ S_0\cdot  \bar S_0\cdot \left(-1-3\Phi -3\bar \Phi + h(S, \bar S) \right) \right]_D
+\left[ S_0^3 \left( S \cdot \left(g(S)+3\Phi \right ) +\lambda \right) \right] _F\,.
\label{or}
\end{equation}
This defines the K{\"a}hler potential and superpotential (see e.g. (17.67)
in \cite{Freedman:2012zz}) as
\begin{equation}
  K(\Phi ,S,\bar \Phi ,\bar S)=-3\log\left( \ft13+\Phi +\bar \Phi -\ft13\, h(S, \bar S) \right)\,,\qquad W(\Phi ,S)=S \cdot \left(g(S)+3\Phi \right ) +\lambda\,.
 \label{KandWgeneral}
\end{equation}

Now we observe that the chiral function $g(S)$ can be removed by defining
\begin{equation}
\Phi'= \Phi +\ft13 g(S)\, , \qquad h'(S, \bar S) = h(S, \bar S) + g(S) + \bar g(\bar S)\,.
\label{defphihprime}
\end{equation}
We finally obtain, according to  \cite{Cecotti:1987sa}
\begin{equation}
{\cal L}_{\rm dual}= \left[ S_0\cdot  \bar S_0\cdot \left(- 1-3\Phi' -3\bar \Phi' + h'(S, \bar S)\right) \right] _D + \left[ S_0^3\cdot
 ( \lambda +3 S \cdot \Phi'  )\right]_F\,.
\label{du}
\end{equation}
We notice that the \K\, potential in \rf{du} is of the no-scale type
\cite{noscale}, while the superpotential is not, because of the $S\cdot
\Phi'$ term. In fact, the no-scale model would give rise to a massless
mode, which is not present in the action \rf{du}.

A model  with no-scale K{\"a}hler potential, and the superpotential as in
(\ref{du}) with $\lambda=0$ has been constructed in
\cite{Kallosh:2013lkr}. It reproduces the bosonic $R+R^2$ model without
unstable directions. In the $R+R^2$ supergravity theory, the
corresponding expression of the $h$ function in (\ref{Sergio})  is
\begin{equation}
S_0 \cdot \bar S_0\cdot h\left({{\cal R}\over S_0}, {\bar {\cal R}\over \bar S_0}\right)
 =  S_0 \cdot \bar S_0 + 3{\cal R}\cdot \bar {\cal R} - \zeta \cdot \frac{({\cal R}\cdot \bar {\cal R})^2}{S_0 \cdot \bar S_0}
 \,,\qquad W\left({{\cal R}\over S_0}\right)=-3{{\cal R}\over S_0}\,.
 \label{exprh}
\end{equation}
The corresponding K{\"a}hler potential and superpotential of the dual theory
are \cite{Kallosh:2013lkr}
\begin{equation}
  K= -3\log\left( \Phi +\bar \Phi -S\cdot \bar S+\ft13\zeta\cdot  (S\cdot \bar S)^2\right)\,,\qquad
  W=3\, S\cdot(\Phi -1)\,,
 \label{KWdual}
\end{equation}
where $\Phi$ is the inflaton multiplet and $S$ is a goldstino multiplet.
More general models in this class corresponding to an action
(\ref{Sergio}) can have extra sgoldstino-dependent terms: in the
superpotential one can add $W(S)$ and in the \K\, potential one can add
some additional terms in $h(S, \bar S)$. If these more general terms do
not destroy the property of the model that $S=0$ is a minimum of the
potential, these more general models will have the same cosmological
observables as the ones in \cite{Kallosh:2013lkr}. This adds additional
features to the cosmological attractors described in
\cite{Kallosh:2013hoa}. Modified non-scale models resembling the
Starobinsky model have also been investigated \cite{EllisStarobinsky}.

The equivalence between \rf{or} and \rf{du} can also be understood by
writing in (\ref{Sergio})
\begin{equation}
W \left({{\cal R}\over S_0}\right)= {{\cal R}\over S_0} \cdot g \Big ({{\cal R}\over S_0}\Big ) +\lambda\,,
\end{equation}
and using once more the lemma (\ref{thmDtoF}) we obtain
\begin{equation}
\left[ {{\cal R}\over S_0} \cdot g \Big ({{\cal R}\over S_0}\Big ) \cdot S_0^3\right] _F=\left[  \Big[ g \Big ({{\cal R}\over S_0}\Big )+ \bar g \Big ({\bar {\cal R}\over \bar S_0}\Big )\Big ]\cdot S_0\cdot  \bar
S_0\right] _D\,,
\end{equation}
which can then be absorbed in a redefinition as in (\ref{defphihprime}).
This shows that the $F({\cal R})= {\cal R} \cdot g({\cal R})$ term
is completely irrelevant
when a non-trivial $h$-function with $h_{S \bar S}\neq  0$ is
introduced.

\subsection{The dual of ``\texorpdfstring{$F(R)$}{F(R)} supergravity'' revisited}
We can now retrieve as a particular case of the dual theory the ``$F(R)$
supergravity''  \cite{Ketov:2009sq,Ketov:2010qz} when $h_{S\bar S}=0$.
In this case $h'=g(S) +\bar g(\bar S)$ so that (\ref{du}) becomes
\begin{equation}
{\cal L}_{\rm dual}= \left[ S_0\cdot  \bar S_0\cdot \left(-1-3\Phi' -3\bar \Phi' + g(S) + \bar g( \bar S) \right)\right] _D
+ \left[ S_0^3\cdot  ( \lambda +3\, S \cdot \Phi'  )\right]_F\,.
\label{du1}
\end{equation}
By defining $\Phi = \Phi'-\ft13 g(S)$ we get
\begin{equation}
{\cal L}_{\rm dual}=  \left[ S_0 \cdot \bar S_0\cdot (-1-3\Phi -3\bar \Phi )\right] _D +  \left[S_0^3\cdot  ( \lambda + S \cdot (3\Phi+ g(S) ) \right] _F\,.
\label{du2}
\end{equation}
Since $S$ is not dynamical, we can now integrate over $S$ so we obtain
the superpotential term $\lambda + w(\Phi)$ given by the Legendre
transform
\begin{equation}
 w(\Phi)=\left. -S^2\cdot  g'(S) \right|_{(S\cdot g(S))'=-3\Phi}\,.
\end{equation}
The final Lagrangian is
\begin{equation}
{\cal L}_{\rm dual}=  \left[ S_0\cdot  \bar S_0\cdot (-1-3\Phi -3\bar \Phi )\right] _D +  \left[S_0^3\cdot  \left( \lambda +w(\Phi )
\right)\right] _F\,.
\label{Lag}
\end{equation}

The Lagrangian \rf{Lag} is dual to ``$F(R)$ supergravity''
\cite{Ketov:2009sq,Ketov:2010qz}, where only the $X$ auxiliary field
becomes propagating and not the $R+R^2$ scalaron mode  and not the
$\partial_\mu A^\mu$ auxiliary scalar, see Sec. \ref{ss:compFR}. In this
situation only the vector field $A_\mu$ is auxiliary and satisfies an
algebraic equation.

\section{The structure of ``\texorpdfstring{$F(R)$}{F(R)} Supergravity''}
\label{ss:compFR}

\subsection{Review of ``$F(R)$ Supergravity'' and cosmology}

The theories that have been called ``$F(R)$ Supergravity'' in
\cite{Ketov:2009sq,Ketov:2010qz} are defined in superspace as
\begin{equation}
  \int \rmd^4x\, \rmd^2\theta \,  {\cal E} \cdot  F({\cal R}(x, \theta)) + \hc\,,
 \label{FRsuperspace}
\end{equation}
where ${\cal E}$ is the chiral superspace density, and ${\cal R}(x, \theta)$ is a chiral
superfield containing the spacetime curvature $R(x)$ in the $\theta^2$
part:
\begin{equation}
  {\cal R}(x, \theta)= X+ \ldots + \theta^2\cdot  R(x)+\ldots \,.
 \label{calRsuperfield}
\end{equation}

This action does not contain the bosonic terms $R+\alpha R^2$. Since in $N=1$ supersymmetry $\theta^{2n}=0$ for $n>1$,
integration of $F({\cal R})$  over $\rmd^2 \theta$ does not produce a non-linear dependence on space-time curvature
$R(x)$.\footnote{For example we may look at the $  {\cal R}^2$ term in $F({\cal R}(x, \theta))$.
The bosonic term $R^2(x)$ there comes with $\theta^4$, which vanishes for $N=1$ supersymmetry since there are only 2 components $\theta_\alpha$.
}
However, if the first component $X$ in (\ref{calRsuperfield}) would be an auxiliary
field, its elimination could produce $R^2$ or higher powers.

For example, with the choice made by Ketov and Starobinsky
\cite{Ketov:2010qz}
\begin{equation}
  F({\cal R})= {1\over 2} f_1 {\cal R}+ {1\over 2} f_2 {\cal R}^2 + {1\over 6} f_3 {\cal R}^3\,,
 \label{FtoR3}
\end{equation}
 the supersymmetric action \rf{FRsuperspace} according to \cite{Ketov:2009sq,Ketov:2010qz} has the  bosonic part
\begin{equation}
{\cal L}_{\rm bos}=\sqrt{g}\cdot  \Re F'(\bar X)\cdot  \left(\ft23 \, R + 8 X\cdot \bar X\right) + 6\sqrt{g}\cdot \Re \left( X \cdot F(\bar X)\right) \,,
\label{bos}
\end{equation}
where $X$ is the complex auxiliary scalar of the supergravity multiplet.
Since the equation of motion for the auxiliary field $X$ following from
\rf{bos} is {\it algebraic}, one solves $X(x)$ in terms of $R(x)$ and in
approximation of high curvature one produces the non-linear in $R(x)$
action
 of the form
\begin{equation}
  {\cal L}= {1\over 2}\sqrt{g}\cdot  \left(R+ {R^2\over 6 M^2} + {3 {\sqrt{ 105}}\over 100} {R^{3/2}\over m}\right)\,,
\label{KetovBos}
\end{equation}
where $M, m$ depend on $f_i$ in  \rf{FtoR3}. Such actions have been used
by cosmologists, see for example \cite{Watanabe:2013lwa} for
applications, such as reheating and non-Gaussianity in supergravity
$R^2$-inflation.

However, we will show in Sec. \ref{ss:actcomp} that the action
(\ref{KetovBos}) is unrelated to the supersymmetric expression in
(\ref{FRsuperspace}), because the field $X$ appears with derivatives.
The  supersymmetric action \rf{FRsuperspace}  in fact has some extra
terms beyond the ones in \rf{bos} depending on $A_\mu$,  an auxiliary
field of supergravity. Rescaling\footnote{The variable $X$ of Sec.
\ref{ss:actcomp} is replaced here by $X/\sqrt{3}$ and $F$ by
$4F/(3\sqrt{3})$.} the result of Sec. \ref{ss:actcomp}  reads
\begin{eqnarray}
  {\cal  L}_{\rm bos}&=&\sqrt{g}\cdot  \Re  F'(\bar X) \cdot \left(\ft23R
  +8\,X\cdot \bar X\right) -12 \sqrt{g}\cdot  \Re\left( X \cdot F(\bar X)\right)\nonumber\\
  &&+4 \sqrt{g}\cdot\Re F'(\bar X)\cdot    A^\mu \cdot  A _\mu -4\sqrt{g}\cdot  A^\mu \cdot   \partial _\mu \Im F'(\bar X) \,.
 \label{Lresultrescaled}
\end{eqnarray}
If $F''(\bar X)=0$ one finds that $A_\mu=0$, however for $F''(\bar
X)\neq 0$ one finds that upon elimination of $A^\mu$ on its equations of
motion, there are extra terms in the bosonic action. For example, in
case of \rf{FtoR3} the action includes terms proportional to (see
complete expression below)
\begin{equation}
  f_2\,   \partial
_\mu X \cdot \partial
^\mu \bar  X +\ldots\,.
 \label{f2example}
\end{equation}
Therefore the complete bosonic part of the supersymmetric action
(\ref{FRsuperspace}) does not lead to an algebraic equation of motion
for $X$, this field is not auxiliary anymore, it is propagating, and in
fact is the only propagating complex scalar degree of freedom.

\subsection{The action in components}
\label{ss:actcomp}

We use the superconformal methods for $N=1$ supergravity with a chiral
compensating multiplet $S_0$, which is reviewed in
\cite{Freedman:2012zz}. The multiplet is identified by its first
component, and we use therefore the same name $S_0$ for this complex
scalar with Weyl weight~1. Using this compensating multiplet implies
that in usual supergravity (without higher curvature terms) we have as
auxiliary fields: the vector field of the Weyl multiplet, $A_a$, and the
complex scalar that is the highest component of $S_0$. That scalar (in
fact its complex conjugate) is proportional to the lowest component of
the kinetic multiplet $\Sigma (S_0)$ (see Appendix \ref{app:elemConf}),
or thus also proportional to the lowest component of ${\cal R}$ defined
in (\ref{defRchiral}). The proportionality is always with factors of
$S_0$, which in a gauge fixing as in (16.40) in
\cite{Freedman:2012zz},\footnote{Such high equation numbers below refer
always to this book.} implies a proportionality with powers of the
gravitational coupling constant $\kappa $. We will parametrize it as the
lowest component of the chiral multiplet that appears in (\ref{Sergio}):
\begin{equation}
  \bar X=\frac{{\cal R}}{S_0}= \frac{\Sigma (S_0)}{S_0^2}\,.
 \label{defX}
\end{equation}
$\bar X$ has Weyl weight~0. We will thus construct
\begin{equation}
  {\cal L}=(-\ft12)\left[ S_0^3 \cdot F(\bar X)\right] _F\,,
 \label{LfromFf}
\end{equation}
which is, up to normalization, the same as (\ref{FRsuperspace}), or as
(\ref{Sergio}) with $W(\bar X)= \ft12 [\bar X - F(\bar X)]$ and $h=0$.
In order to evaluate (\ref{LfromFf}), we need the highest components of
the various multiplets. Using the definition (\ref{defX}) and (16.36),
which gives the highest component of $\Sigma (Z)$ for any chiral
multiplet $Z$, the various multiplets have the following highest
components:
\begin{equation}
  S_0\,\rightarrow\, S_0^2\cdot  X\,,\qquad \Sigma (S_0)\,\rightarrow\,\bbox^C\bar S_0\,,\qquad
  \bar X\,\rightarrow \, S_0^{-2}\cdot \bbox^C\bar S_0-2\,S_0\cdot X\cdot  \bar X\,.
 \label{Fcomponents}
\end{equation}
Therefore, the bosonic part of (\ref{LfromFf}) is
\begin{equation}
 {\cal  L}_{\rm bos}=-\ft12 \sqrt{g}\cdot \left\{ 3S_0^4\cdot  X \cdot F(\bar X)+ S_0^3\cdot  F'(\bar X)\cdot \left(S_0^{-2}\cdot \bbox^C\bar S_0-2\,S_0\cdot X\cdot  \bar X\right)\right\} +\hc\,.
 \label{LderivF}
\end{equation}

We will now use a gauge-fixing of the dilatations, where $S_0$ is a
constant. Then the superconformal d'Alembertian gets the simple form as
in (16.42) We can use here (16.37) for the $\bbox^C \bar S_0$:
\begin{equation}
 \bbox^C \bar S_0 =\bar S_0\cdot \left[ \rmi e^{a\mu}\cdot \left(
\partial_\mu  A _a +\omega_{\mu\,ab} \cdot A^b \right)+2f_\mu {}^\mu - A^a \cdot A _a\right]  +\mbox{fermionic}\,,
 \label{boxCZ}
\end{equation}
and $f_\mu {}^\mu =-\ft1{12}R$. With the gauge fixing $S_0=\bar
S_0=\sqrt{3}$, we obtain
\begin{eqnarray}
 {\cal  L}_{\rm bos}&=&-\ft12\sqrt{g}\cdot\left\{  27\, X \cdot F(\bar X)\right.  \label{LBgf}\\
 && \left.+3\,F'(\bar X)\cdot \left[ \rmi e^{a\mu}\left(
\partial_\mu  A _a +\omega_{\mu\,ab}\cdot  A^b  \right)-\ft16R- A^a \cdot A _a  -6X\cdot \bar X \right]\right\}+\hc \,.\nonumber
\end{eqnarray}
Note that for the case $W(\bar X)=0$, i.e. $F(\bar X)= \bar X$, the term
with the derivative of $A _a$ is a total derivative, and this reproduces
pure supergravity as in (16.44), now rewritten in terms of $X$. In that
case also $X$ is auxiliary.

After adding a total derivative, we can write
\begin{eqnarray}
  {\cal  L}_{\rm bos}&=&-\ft12\sqrt{g}\cdot\left\{ 27\, X\cdot  F(\bar X)- 3\rmi \, A^\mu \cdot  \partial _\mu F'(\bar X)\right.\nonumber\\
   &&\left.+F'(\bar X)\cdot \left[-\ft12R
  - 3A^\mu \cdot  A _\mu    -18X\cdot \bar X\right]\right\}+\hc\,.
 \label{LBKetov2}
\end{eqnarray}
The field equation that follows from (\ref{LBKetov2}) determines
\begin{equation}
  A_\mu =\frac{\partial _\mu \Im\left(F'(\bar X)\right) }{2\Re\left(F'(\bar X)\right)} \,.
 \label{AmusolndX}
\end{equation}
The action after inserting this value is
\begin{eqnarray}
  {\cal  L}_{\rm bos}&=&\sqrt{g}\cdot\left\{ -27\, \Re\left( X \cdot F(\bar X)\right) +\ft12 \left(\Re  F'(\bar X)\right) \cdot \left(R
  +36\,X\cdot \bar X\right)\right.\nonumber\\
  && \left.-\frac{3}{4} \frac{\left[ \partial _\mu \Im\left(F'(\bar X)\right)\right] ^2 }{\Re\left(F'(\bar X)\right)} \right\} \,.
 \label{LelimA}
\end{eqnarray}

Hence, the implications of $F(\bar X)$ are as follows: we have seen
above that a linear term in $\bar X$ (with minus sign for the signature
of the gravity term) produces pure supergravity. Adding a constant term
produces a cosmological term (anti-de Sitter). When $F''(\bar X)\neq 0$,
there are two real scalar physical fields: the imaginary part of
$F'(\bar X)$ has its kinetic term explicitly in the last term. The real
part of $F'(\bar X)$ is coupled to $R$. The action in (\ref{LelimA}) is
presented in a Jordan frame, and the transformation to Einstein frame
transforms this term as
\begin{equation}
  {\cal L}_{\rm J}= \frac12 \sqrt{g}\cdot \Re  F'(\bar X) \cdot R\ \rightarrow
  {\cal L}_{\rm E} =   \sqrt{g}\cdot \left[\frac12 R -\frac34\left( \frac{\partial _\mu \Re F'(\bar X)}{\Re F'(\bar X)}\right) ^2\right] \,.
 \label{JordanEinstein}
\end{equation}
So this theory describes a propagating complex scalar ($X=S+\rmi P$) and
is not an $R+R^2$ theory.

\newpage

\section{Discussion}
\label{ss:discussion}

In this paper we have revisited the supersymmetric completion of $R+R^2$
supergravity in the old minimal formulation. The manifestly
superconformal action is given in \rf{Sergio}. It is defined by a real
function $h\left({{\cal R}\over S_0}, {\bar {\cal R}\over \bar
S_0}\right)$ and a holomorphic function $W\left ( {\bar {\cal R}\over
\bar S_0}\right)$. Here $S_0$ is a chiral conformon supermultiplet and
${\cal R}$ is the curvature supermultiplet.

We have elucidated its degrees of freedom and the corresponding
properties of the dual theory, in agreement with the original results of
\cite{Cecotti:1987sa}. We stress the fact that any supersymmetric theory
of $R+R^2$ gravity has to contain two chiral multiplets with some
universal interactions. The first multiplet, is the inflaton $\Phi$, the
second one is the goldstino $S$, according to their role during
inflation.

We have found a new form of the \K\,-type symmetry between
$h\left({{\cal R}\over S_0}, {\bar {\cal R}\over \bar S_0}\right)$ and
$W\left ( {\bar {\cal R}\over \bar S_0}\right)$ at the level of the
superconformal theory, before the extra symmetries absent in Poincar{\'e}
supergravity are fixed. Namely, we have shown that there is a
possibility to absorb the part of the superpotential of the form $W(S) =
S\cdot  g(S)$ into the modification of the real function $h(S, \bar
S)\rightarrow h(S, \bar S) +g(S) + \bar g(\bar S)$.

We described the relation of this class of theories with the
cosmological model of inflation in \cite{Kallosh:2013lkr}, where the
corresponding \K\, potential and superpotential of the ordinary
two-derivative supergravity are given in \rf {KWdual}.  During inflation
only the real scalar of the inflaton multiplet (the inflaton) is
evolving; the 3 other scalars are stabilized in this model. The most
general models  dual to a supersymmetric completion of $R+R^2$ in
\rf{Sergio} may add to \cite{Kallosh:2013lkr} some arbitrary
superpotentials $W(S)$,  depending only on the goldstino multiplet, as
well as more general goldstino depending terms $A(S, \bar S)$  in the
\K\, potential of the form $ K_A= -3\log\left( \Phi +\bar \Phi -S\cdot
\bar S+\ft13\zeta\cdot  (S\cdot \bar S)^2 + A(S, \bar S)\right)$. If in
presence of these new terms $W(S)$ and $A(S, \bar S)$ the generalized
models still have a stable minimum at the vanishing sgoldstino, we find
a new attractor property of these models in addition to the one
described in \cite{Kallosh:2013hoa}: the inflaton potential and,
consequently, the cosmological observables during inflation do not
depend on the choice of $W(S)$ and $A(S, \bar S)$. An example of such a
deformation is a superpotential $gS^3$ with a sufficiently small $g$.

We have then revisited the so-called ``$F(R)$ supergravity''
\cite{Ketov:2009sq,Ketov:2010qz}, where only one chiral multiplet
occurs. We clarified the physical degrees of freedom of the gravity
theory and its dual counterpart and showed that the bosonic action of
``$F(R)$ supergravity'' \cite{Ketov:2009sq,Ketov:2010qz} is linear in
the space-time curvature $R(x)$ and does not have non-linear terms in
the curvature. In particular, it does not contain $R^2(x)$ term required
for inflation in this model.

It is interesting to observe that in the cosmological models based on
supersymmetry of which $R+R^2$ supergravity is a particular realization,
three chiral multiplets come to play a role: in addition to an inflaton
multiplet $\Phi$ and the goldstino multiplet $S$, also the conformon
(the conformal compensator multiplet $S_0$) has an important role. The
later helps in the use of some hidden symmetries that are remnants of
the full underlying superconformal formulation of the theory.

\section*{Acknowledgement}
We are grateful to S. Cecotti,  A. Linde, and M. Porrati   for
stimulating discussions.

R.K. is supported by Stanford Institute for Theoretical Physics (SITP),
the NSF Grant No. 0756174 and the John Templeton foundation grant
`Quantum Gravity Frontiers'. S.F.   is supported by ERC Advanced
Investigator Grant n. 226455 {\em Supersymmetry, Quantum Gravity and
Gauge Fields (Superfields)}. A.V.P. is supported in part by the FWO -
Vlaanderen, Project No. G.0651.11, and in part by the Interuniversity
Attraction Poles Programme initiated by the Belgian Science Policy
(P7/37). S.F and A.V.P. are grateful to SITP  for the hospitality at
Stanford where this work was finalized.

\appendix
\addtocontents{toc}{\protect\setcounter{tocdepth}{1}}
\section{Elements of conformal tensor calculus}
 \label{app:elemConf}

The chiral projection is in superspace an operation $\bar D^2$, applied
to the complex conjugate superfield. We define on a chiral multiplet $S$
with (Weyl, chiral) weight (see Table 17.1) $(w,c)=(1,1)$ the map to
another chiral multiplet by
\begin{equation}
  \Sigma (S)= T(\bar S)\ :\ S\ (w,c)=(1,1)\rightarrow \Sigma (S)=T(\bar S)\ (w,c)=(2,2)\,.
 \label{SigmaT}
\end{equation}
This is the map that is explicitly given in (16.36), which associates to
the chiral multiplet $S=(X,P_L\Omega ,F)$ the multiplet starting with
$\bar F$.

For any two chiral multiplets $\Lambda $ (with $w=0$) and $Z$ (with
$w=1$) we have the lemma
\begin{equation}
 \left[Z\cdot \bar Z\cdot (\Lambda +\bar \Lambda )\right]_D=  \left[\Lambda\cdot  Z\cdot  \Sigma (Z)\right]_F\,.
 \label{thmDtoF}
\end{equation}
We show this by checking the bosonic parts. The left-hand side is a
function of chiral multiplets, and we can use (17.19):
\begin{eqnarray}
 \left[Z\cdot \bar Z\cdot (\Lambda +\bar \Lambda )\right]_D & = &\sqrt{g}\cdot \left(  -{\cal D}_\mu (\Lambda Z)\cdot {\cal D}^\mu \bar Z
 + \Lambda _F\cdot  Z\cdot \bar F+\Lambda\cdot  F\cdot \bar F\right.\nonumber\\
 &&\phantom{\sqrt{g}\cdot A}\left.-\ft16\Lambda\cdot  Z\cdot \bar Z\cdot  R(\omega )\right) +\hc \nonumber\\
   & = & \sqrt{g}\cdot \left( (\Lambda\cdot  Z)\cdot \Box^C \bar Z
 + \Lambda _F\cdot  Z\cdot \bar F+\Lambda\cdot  F\cdot \bar F\right) +\hc\,,
 \label{proofDtoF}
\end{eqnarray}
where $\Lambda _F$ is the $F$ component of $\Lambda $. From (16.36) we
see that $\Box^C \bar Z$ is the $F$ component of $\Sigma (Z)$. Thus the
three terms are the 3 ways in which to get the $F$ component indicated
in the right-hand side of (\ref{thmDtoF}), which proves the lemma.

%

\newpage

\providecommand{\href}[2]{#2}\begingroup\raggedright\endgroup

\end{document}